\documentclass[twocolumn,pra,showpacs,nofootinbib,superscriptaddress]{revtex4}

\usepackage{dcolumn,bm,graphicx,amsmath}
\usepackage
{hyperref}

\newcommand{\eref}[1]{(\ref{#1})}
\newcommand{\fref}[1]{Fig.~\ref{#1}}

\newcommand{\sref}[1]{Sec. \ref{#1}}
\newcommand{\Eref}[1]{Eq.~(\ref{#1})}
\newcommand{\tref}[1]{Table~\ref{#1}}
\def\veps{\varepsilon}

\begin{document}
\preprint{UNR Jan 2005-\today }
\title{ Atomic CP-violating polarizability}

\author{Boris Ravaine}
\affiliation{Department of Physics, University of Nevada, Reno,
Nevada 89557}
\author{M. G. Kozlov}
\email{mgk@MF1309.spb.edu}
\affiliation{Petersburg Nuclear Physics Institute, Gatchina
188300, Russia}
\author{Andrei Derevianko}
\email{andrei@unr.edu}
\affiliation{Department of Physics, University of Nevada, Reno,
Nevada 89557}

\date{\today}

\begin{abstract}
Searches for  CP violating effects in atoms and molecules
provide important constrains on competing extensions to the
standard model of elementary particles. In particular, CP violation in an atom
leads to the CP-odd (T,P-odd) polarizability $\beta^\mathrm{CP}$:
a magnetic moment $\mu^\mathrm{CP}$
is induced by an electric field $\mathcal{E}_0$ applied to an atom,
$\mu^\mathrm{CP} = \beta^\mathrm{CP} \mathcal{E}_0 $.
We estimate
the CP-violating polarizability for rare-gas (diamagnetic) atoms He through Rn.
We relate $\beta^\mathrm{CP}$  to the permanent electric dipole moment (EDM)
of the electron and to the scalar constant of the CP-odd electron-nucleus interaction.
The analysis is carried out using the third-order perturbation theory and the Dirac-Hartree-Fock
formalism. We find that, as a function of nuclear charge $Z$, $\beta^\mathrm{CP}$  scales steeply
 as $Z^5 R(Z)$, where slowly-varying $R(Z)$ is a relativistic enhancement factor. Finally, we
evaluate a feasibility of setting a limit on electron EDM by measuring CP-violating magnetization
of liquid Xe.
We find that such an experiment could provide competitive bounds on electron EDM only
if the present level of experimental sensitivity to ultra-weak magnetic fields
[Kominis {\em et al.}, Nature {\bf 422}, 596 (2003)] is
improved by several orders of magnitude.
\end{abstract}

\pacs{11.30.Er,32.10.Dk,31.30.Jv}

\maketitle

\section{Introduction}

The existence of a permanent electric dipole moment (EDM) of a particle
simultaneously violates two discrete symmetries: parity (P) and
time reversal (T). By the virtue of the CPT-theorem, the
T-violation would imply CP-violation~\cite{KL97,BS00}.
While no EDMs have been found so far, most supersymmetric
extensions of the Standard Model of elementary particles
predict electron EDMs, $d_e$, that are within a reach of planned and
on-going experimental searches. Here we investigate a related
T-odd, P-odd  quantity --- CP-violating polarizability, $\beta^\mathrm{CP}$,  introduced
recently by \citet{Bar04}. For a diamagnetic atom, a non-vanishing $\beta^\mathrm{CP}$
could provide an unambiguous signature of the electron EDM or other CP-violating
mechanisms. Here we relate $\beta^\mathrm{CP}$ to
$d_e$ via {\em ab initio} relativistic calculations for closed-shell
atoms. We also relate $\beta^\mathrm{CP}$ to the scalar constant of
the CP-odd electron-nucleus interaction.

An interaction of an atom with external DC electric field
 in the presence of the electron EDM
causes spin polarization in the direction of the
field \cite{Sha68}.
The first attempt to measure corresponding magnetization of the
ferromagnetic crystal was made by Vasiliev and Kolycheva in 1978
\cite{VK78}. According to \citet{Lam01}, modern techniques allow
to improve that old measurement by many orders of magnitude and
reach the sensitivity, which allows to improve present limit on
the electron EDM \cite{RegComSch02}:
\[
d_{e}\left( \mathrm{Tl} \right) < 1.6 \times 10^{-27} e\cdot
\mathrm{cm}.
\]
First results of the new generation of experiments with
ferromagnetic solids were recently reported by \citet{Hun04}.
Characteristic feature of the experiments with macroscopic
magnetization is the dependence of the signal on the density of
atoms. That gives a huge enhancement in sensitivity for a condensed
phase sample.

It is generally assumed that diamagnetic atoms are not useful for
the search of the electron EDM. However, Baryshevsky has recently
pointed out~\cite{Bar04} that CP-violating magnetization would also exists
for a diamagnetic atom. For a spherically-symmetric
atom, the E-field-induced magnetic moment $\mu^{\mathrm{CP}}$  can
be expressed in terms of CP-violating polarizability $\beta^{\mathrm{CP}}$ as
\begin{equation}\label{i1}
\mu^{\mathrm{CP}} = \beta^{\mathrm{CP}} \mathcal{E}_{0},
\end{equation}
where $\mathcal{E}_{0}$ is the strength of the electric field.
This observation opens new experimental possibilities.
For example, one can measure magnetization of liquid
xenon in a strong external electric field. The advantage of the
experiment with diamagnetic liquid in comparison to ferromagnetic
solids is a much lower magnetic noise.

For a diamagnetic (closed-shell) atom the magnetization \eref{i1} appears in the
higher orders of the perturbation theory than for the open-shell atoms.
In this paper we calculate polarizability
$\beta^{\mathrm{CP}}$ for rare-gas atoms He through Rn
using third-order perturbation theory and Dirac-Hartree-Fock (DHF)
formalism.

Further,  we
evaluate a feasibility of setting a limit on electron EDM by measuring CP-violating magnetization
of liquid Xe (LXe).  We consider the
effect of the environment on $\beta^{\mathrm{CP}}$ of Xe atoms in LXe.
We use  a simple cell
model of an atom confined in a spherically-symmetric cavity~\cite{RD04}. In
a non-polar liquid, such a cavity roughly approximates an averaged
interaction with the neighboring atoms. We solve the
DHF equations with proper boundary
conditions at the cavity radius. For LXe, we find that compared to the
CP-odd polarizability of an isolated atom, the resulting
CP-odd polarizability of an atom of LXe is suppressed by
about 65\%.

We find that the CP-violating polarizability exhibits an unusually strong dependence
on the nuclear charge $Z$. Previously,
Sandars \cite{San65,San66} has shown that an atomic enhancement
factor for the electron EDM is of the order of $\alpha^2 Z^3$,
where $\alpha\approx 1/137$ is the fine structure constant. As we demonstrate below, for a
diamagnetic atom, the polarizability $\beta^{\mathrm{CP}}$ vanishes in
the non-relativistic approximation. Because of that it is
suppressed by a factor of $(\alpha Z)^2$. With the Sandars' enhancement
factor this leads to a steep, $Z^5$, scaling of the effect.

Recently there was a renewed interest to the CP-odd weak neutral
current interactions of electrons with nucleons \cite{Jun05}. It
is known that in atomic experiments the electron EDM is
indistinguishable from the scalar CP-odd weak neutral currents
\cite{Khr91,KL97}. Any new limit on the electron EDM from the atomic
experiments will also lead to the improved limit on the scalar
constant of the CP-odd electron-nuclear interaction. Here we
relate  computed $\beta^\mathrm{CP}$ to the scalar constant of the CP-odd
electron-nucleus interaction.

The paper is organized as follows: In Section~\ref{Sec:Formalism}
we derive the third-order expression for the CP-violating polarizability
and use the independent-particle approximation to simplify the
atomic many-body expressions. In Section~\ref{Sec:Results} we
present results of our DHF calculations of $\beta^{\mathrm{CP}}$
for rare-gas atoms and derive the $Z$-scaling of $\beta^{\mathrm{CP}}$.
In Section~\ref{Sec:LXe} we
evaluate a feasibility of setting a limit on electron EDM by measuring CP-violating magnetization
of liquid Xe. Finally, in Section~\ref{Sec:Conclusion} we draw conclusions.
 Unless specified otherwise,
atomic units $|e|=\hbar=m_e\equiv 1$ and Gaussian system for
electro-magnetic equations are used throughout. In these units,
the Bohr magneton is $\mu_B=\alpha/2$ and the unit of magnetic field is
$m_e^2 e^5/\hbar^4 \approx 1.72 \times 10^7 \, \mathrm{Gauss}$.

\section{Formalism}
\label{Sec:Formalism}
In this Section we derive the expression for CP-violating polarizability
within the third-order perturbation theory. Further, we simplify the
derived
expression using the Dirac-Hartree-Fock approximation for atomic
many-body states.

The problem to be solved can be formulated as follows:
What is the induced magnetic moment $\left\langle \bm{\mu}\right\rangle$
of an atom perturbed by
an external electric field $\mathcal{E}_{0}$? It is easy to demonstrate
that if the atomic wavefunctions are the eigenstates of the parity and
time-reversal operators,
the induced magnetic moment vanishes. However, in the presence of the
CP-odd interactions,
$V^\mathrm{CP}$, there appears a tiny E-field-induced magnetic moment.
To emphasize
the essential role of CP-violation in the generation of the
magnetic-moment, we
will use CP superscript with the magnetic moment, $\left\langle
\bm{\mu}^\mathrm{CP}\right\rangle$.
The interaction $V^\mathrm{CP}$ can be due to electron EDM or CP-odd
weak neutral currents,
and we will specify the particular forms of $V^\mathrm{CP}$ in Section~\ref{Sec:mels}.
For a spherically-symmetric system, the induced magnetic moment
will be directed along the applied E-field.

\subsection{Third-order formula for the induced magnetic moment}

We develop the perturbative expansion for the atomic wavefunction
$|\Psi_0\rangle$
in terms of the combined interaction
$W =  V^\mathrm{CP}+V^\mathrm{ext}$. Here $V^\mathrm{ext}$ is
the interaction with the external electric field applied along the
$z$-axis,
$V^\mathrm{ext}  = -D_{z} \, \mathcal{E}_{0}$, $D_z$ being the
z-component of the electric dipole moment
operator.
To estimate the dominant contribution to $\left\langle
\bm{\mu}\right\rangle$,
it is sufficient to truncate the perturbative expansion
for the atomic wavefunction at the second order in $W$,
$|\Psi_0\rangle \approx  |\Psi_{0}^{\left(  0\right)  }\rangle +
 |\Psi_{0}^{\left(  1\right)  }\rangle +|\Psi_{0}^{\left(  2\right)
}\rangle $.
Then the expectation value of the magnetic moment reads
\begin{equation}
\left\langle \bm{\mu}^\mathrm{CP}\right\rangle =
\langle\Psi_{0}^{\left(  1\right)  }|\bm{\mu}|\Psi_{0}^{\left(  1\right)
 }\rangle +
\langle\Psi_{0}^{\left(  0\right)
}|\bm{\mu}|\Psi_{0}^{\left(  2\right)  }\rangle+\langle\Psi_{0}^{\left(
2\right)  }|\bm{\mu}|\Psi_{0}^{\left(  0\right)  }\rangle
 \,.
\end{equation}
To arrive at the above expression we used a simplifying fact that the
magnetic moment
is a P-even operator, while both $|\Psi_{0}^{\left(  0\right)  }\rangle$
and $|\Psi_{0}^{\left(  2\right)  }\rangle$ have parities opposite
to the one of the first-order correction $|\Psi_{0}^{\left(  1\right)
}\rangle$.

The textbook expressions for the first and second-order corrections to
wavefunctions
can be found, for example, in Ref.~\cite{LanLif97}.  With these
expressions,
\begin{eqnarray}
\left\langle \bm{\mu}^\mathrm{CP}\right\rangle  &  =  &
\left\langle \bm{\mu}^\mathrm{CP}\right\rangle_1 +
\left\langle \bm{\mu}^\mathrm{CP}\right\rangle_2 +
\left\langle \bm{\mu}^\mathrm{CP}\right\rangle_3 \,,
\label{Eq:muBreakPT}\\
\left\langle \bm{\mu}^\mathrm{CP}\right\rangle_1 & =  &
2 \sum_{kl}\frac
{V_{0k}^\mathrm{CP}}{E_{0}-E_{k}}
\bm{\mu}_{kl}\frac{V_{l0}^\mathrm{ext}}{E_{0}-E_{l}}\, ,
\label{Eq:muBreakPT1}\\
\left\langle \bm{\mu}^\mathrm{CP}\right\rangle_2 & =  &
2\sum_{kl}
\bm{\mu}_{0k}
\frac{V_{kl}^\mathrm{CP}~V_{l0}^\mathrm{ext}}{\left(
E_{0}-E_{k}\right)  \left(  E_{0}-E_{l}\right)  } \, ,
\label{Eq:muBreakPT2}\\
\left\langle \bm{\mu}^\mathrm{CP}\right\rangle_3 & =  &
  2\sum_{kl}
\bm{\mu}_{0k}
\frac{V_{kl}^\mathrm{ext}\,V_{l0}^\mathrm{CP}}{\left(
E_{0}-E_{k}\right)  \left(
E_{0}-E_{l}\right)  } \, . \label{Eq:muBreakPT3}
\end{eqnarray}
In these formulas, the summations are carried out over the eigenstates
of the atomic Hamiltonian $H_a$,
$H_a|\Psi_{p}^{\left(  0\right)  } \rangle =  E_p |\Psi_{p}^{\left(
0\right)  }\rangle$.
The derived third-order expression can be presented in a more compact
and
symmetrical form using the resolvent operator
$\mathcal{R}= \left(E_{0}-H_a\right)^{-1}$,
\begin{eqnarray}
\lefteqn{ \left\langle \bm{\mu}^\mathrm{CP}\right\rangle =
2 \langle 0 |V^\mathrm{CP}\,\mathcal{R}\,
\bm{\mu}\,\mathcal{R}\,V^\mathrm{ext}|0\rangle+} \label{Eq:MuResolvent}
\\
&& 2\langle
0|\bm{\mu}\,\mathcal{R}\,V^\mathrm{CP}\mathcal{R}\,V^\mathrm{ext}|0\rangle
+2\langle
0|\bm{\mu}\,\mathcal{R}\,V^\mathrm{ext}\,\mathcal{R}\,V^\mathrm{CP}|0\rangle \, .
\nonumber
\end{eqnarray}
The three above contributions differ by permutations of the operators
$\bm{\mu}$, $V^\mathrm{CP}$ and $V^\mathrm{ext}$.

\subsection{Dirac-Hartree-Fock  approximation}
Having derived a general third-order expression for the induced magnetic
moment, Eq.~(\ref{Eq:MuResolvent}),
here we proceed with the atomic-structure part of the evaluation. We
employ the conventional
Hartree-Fock (HF) or independent-particle approximation for that
purpose. In this approach,
the atomic many-body wavefunction is represented by the Slater
determinant
composed of single-particle orbitals. These orbitals are
determined from a set of the HF equations. Using a complete set
of Slater determinants, the contributions to the induced magnetic
moment,
 Eq.(\ref{Eq:muBreakPT1}--\ref{Eq:muBreakPT3}), may be expressed as
\begin{align}
\left\langle \bm{\mu}^\mathrm{CP}\right\rangle_{1,a} &
=
2\sum_{amn}\frac{V_{an}^\mathrm{CP}\bm{\mu}_{nm}\,\,V_{ma}^\mathrm{ext}}
{(\varepsilon_{m}-\varepsilon_{a})\,(\varepsilon_{n}-\varepsilon_{a})}\,
,\label{Eq:muDHF1a} \\
\left\langle \bm{\mu}^\mathrm{CP}\right\rangle_{1,b} &
=
-2\sum_{abm}%
\frac{V_{bm}^\mathrm{CP}\,\bm{\mu}_{ab}\,V_{ma}^\mathrm{ext}}{(\varepsilon
_{m}-\varepsilon_{a})\,(\varepsilon_{m}-\varepsilon_{b})},
\label{Eq:muDHF1b}\\
\left\langle \bm{\mu}^\mathrm{CP}\right\rangle_{2,a} &
= 2\sum_{amn}\frac{\bm{\mu}_{an}V_{nm}^\mathrm{CP}\,V_{ma}^\mathrm{ext}%
}{(\varepsilon_{m}-\varepsilon_{a})\,(\varepsilon_{n}-\varepsilon_{a})}%
\, \, , \label{Eq:muDHF2a}\\
\left\langle \bm{\mu}^\mathrm{CP}\right\rangle_{2,b} &
= -2\sum_{abm}\frac{\bm{\mu}_{bm}V_{ab}^\mathrm{CP}\,V_{ma}^\mathrm{ext}%
}{(\varepsilon_{m}-\varepsilon_{a})\,(\varepsilon_{m}-\varepsilon_{b})},
\label{Eq:muDHF2b}\\
\left\langle \bm{\mu}^\mathrm{CP}\right\rangle_{3,a}
&= 2\sum_{amn}\frac{\bm{\mu}_{an}V_{nm}^\mathrm{ext}\,\,V_{ma}^\mathrm{CP}%
}{(\varepsilon_{m}-\varepsilon_{a})\,(\varepsilon_{n}-\varepsilon_{a})}%
 \, , \label{Eq:muDHF3a}\\
\left\langle \bm{\mu}^\mathrm{CP}\right\rangle_{3,b} &
=
-2\sum_{abm}\frac{\bm{\mu}_{bm}V_{ab}^\mathrm{ext}\,\,V_{ma}^\mathrm{CP}%
}{(\varepsilon_{m}-\varepsilon_{a})\,(\varepsilon_{m}-\varepsilon_{b})}
\, .
\label{Eq:muDHF3b}
\end{align}
Here indexes $a$ and $b$ run over single-particle orbitals occupied in
$|\Psi_0\rangle $, indexes $m$ and $n$ run over virtual orbitals,
and $\varepsilon_{i}$ are the energies of the HF orbitals.

It is well known that the relativistic effects are essential
for the non-vanishing contributions to energy levels due to EDMs (Schiff
theorem).
Moreover, in Section~\ref{Sec:Z5scaling},
we will demonstrate that the relativism enters into the calculations
of CP-violating polarizability in the enhanced fashion: one also needs
to incorporate relativistic corrections to electric- and magnetic-dipole
matrix elements and
energies entering Eq.(\ref{Eq:muBreakPT1}--\ref{Eq:muBreakPT3}).
We include the relativistic effects by directly solving
Dirac-Hartree-Fock (DHF) equations
\begin{equation}
\left( c (\bm{\alpha} \cdot \bm{p}) + \beta c^2 +
V_\mathrm{nuc} + V_\mathrm{DHF} \right) u_i(\bm{r}) =
\varepsilon_i u_i(\bm{r}) \, ,
\end{equation}
where $V_\mathrm{nuc}$ is a potential of the Coulomb interaction
with a finite-size nucleus
and $V_\mathrm{DHF}$ is non-local self-consistent DHF potential.

\subsection{Matrix elements}
\label{Sec:mels}
We use the following ansatz for the Dirac bi-spinor
\begin{equation}
u_{n\kappa m}(\bm{r})= \frac{1}{r}\left(
\begin{array}
[c]{c}%
iP_{n\kappa}(r)\ \Omega_{\kappa m}(\hat{r})\\
Q_{n\kappa}(r)\ \Omega_{-\kappa m}(\hat{r})
\end{array}\,
\right) \, , \label{Eq:BiSpinorSpher}%
\end{equation}
where $P$ and $Q$ are the large and small radial components respectively
and $ \Omega$ is the spherical spinor. The angular quantum number
$\kappa= (l-j)\left(2j+1\right)$.

In particular, the reduced matrix elements of the magnetic-dipole and
electric-dipole moment operators between two bi-spinors
are given by
\begin{eqnarray}
\langle a||\mu||b\rangle &= & \frac{1}{2}\left(  \kappa
_{a}+\kappa_{b}\right)  \,\langle-\kappa_{a}||C_{1}||\kappa_{b}\rangle
\times
\label{Eq:mu_ME}\\
&&\int_{0}^{\infty}r~dr\{P_{a}\left(  r\right)  Q_{b}\left(  r\right)
+Q_{a}\left(  r\right)  P_{b}\left(  r\right)  \} \, , \nonumber \\
\langle a||D||b\rangle &= &
-\langle\kappa_{a}||C_{1}||\kappa_{b}\rangle \times
\label{Eq:D_ME}\\
&&\int_{0}^{\infty}r~dr\{P_{a}\left(  r\right)  P_{b}\left(  r\right)
+Q_{a}\left(  r\right)  Q_{b}\left(  r\right)  \} \, , \nonumber
\end{eqnarray}
$C_1(\hat{r})$ being normalized spherical harmonic.

At this point we would like to specify particular forms for the CP-odd
interaction $V^\mathrm{CP}$. We will distinguish between
the electron EDM coupling $V^\mathrm{CP\!,EDM}$ and
weak neutral-current (NC) interactions $V^\mathrm{CP\!,NC}$.
The EDM interaction of an electron with an electric field
$\mathcal{E}_\mathrm{int}$ can be written in four-component Dirac
notation as ~\cite{Khr91}:
\begin{equation}
V^\mathrm{CP\!,EDM}= 2 d_{e}\left( \begin{array}{cc}  0 & 0 \\
  0 & \bm{\sigma} \cdot \mathcal{E}_\mathrm{int}
\end{array} \right) \label{Eq:H_{e}} .
\end{equation}
The matrix element of this interaction
is given by
\begin{equation}
V^\mathrm{CP,EDM}_{ab} = d_{e} \left\{  2Z\int_{0}^{\infty}\frac{dr}%
{r^{2}}Q_{a}\left(  r\right)  Q_{b}\left(  r\right)  \right\}  \delta
_{\kappa_{a},-\kappa_{b}}\delta_{m_{a},m_{b}} \, ,
\label{Eq:EDM}
\end{equation}
where we assumed that the dominant contribution is accumulated
close to the nucleus (of charge $Z$) so that
$\mathcal{E}_\mathrm{int}$ can be approximated by the nuclear
field. The selection rules with respect to angular quantum numbers
$m$ and $\kappa$ arise because $V^\mathrm{CP}$ is a pseudoscalar.

Recently there was a renewed interest to CP-odd weak neutral
current interactions of electrons with nucleons \cite{Jun05}.
It is known that in atomic experiments EDM of the electron is
indistinguishable from the scalar CP-odd weak neutral currents
\cite{Khr91}:
\begin{align}
V^\mathrm{CP\!,NC}
&= \textrm{i}\frac{G_\textrm{F}}{\sqrt{2}}
(Zk_1^p+Nk_1^n)\gamma_0\gamma_5 \rho(\bm{r}),
\nonumber\\
&\equiv \textrm{i}\frac{G_\textrm{F}Z}{\sqrt{2}}
k_1^\mathrm{nuc}\gamma_0\gamma_5 \rho(\bm{r}),
\label{Eq:NC}
\end{align}
where $G_\textrm{F}= 2.2225\times 10^{-14}$~a.u. is the Fermi
constant, $k_1^{p,n}$ are dimensionless constants of the scalar
$P,T$-odd weak neutral currents for proton and neutron
($k_1^\mathrm{nuc}\equiv k_1^p+\frac{N}{Z}k_1^n$). Further, $Z$ and $N$ are
the numbers of protons and neutrons in the nucleus, $\gamma_{0,5}$
are Dirac matrices, and $\rho(\bm{r})$
is the nuclear density.

\section{Results for rare-gas atoms}
\label{Sec:Results}
The derived HF expressions hold for any atomic or molecular system with
a state  composed from a single Slater determinant. Below we will carry
out calculations for the rare-gas atoms He through Rn. These
closed-shell atoms have a
 $^1\!S_0$   ground state and, due to the spherical symmetry, the
CP-violating
polarizability is a scalar quantity, i.e., the induced magnetic moment
is parallel to the applied electric field. The intermediate many-body
states
in Eq.~(\ref{Eq:muBreakPT1}--\ref{Eq:muBreakPT3}) are particle-hole
excitations,
with the total  angular momenta of $J= 0$ or $J= 1$, depending on the
multipolarity
of the involved operator.

To carry out the numerical evaluation, we solved the DHF equations in
the cavity using a B-spline
basis set technique by \citet{JohBluSap88}. The resulting set of
basis functions is finite and can be considered as numerically complete.
In a typical calculation
we used a set of basis functions expanded over 100 B-splines.
An additional peculiarity related to the Dirac equation is
an appearance of negative energy states ($\varepsilon_m < -m_e c^2$)
in the summation over intermediate states in
Eq.~(\ref{Eq:muDHF1a})--(\ref{Eq:muDHF3b}). In our calculations
we used the so-called length-form of the electric-dipole operator, Eq.~(\ref{Eq:D_ME})
and we found the contribution of negative-energy-state to be insignificant.

\begin{center}
\begin{table}[h]
\begin{tabular}{lrdd}
\hline\hline \multicolumn{1}{c}{Atom} & \multicolumn{1}{r}{Z} &
\multicolumn{1}{r}{$\beta^\mathrm{CP}/d_{e}$} &
\multicolumn{1}{r}{$\beta^\mathrm{CP}/k_1^\mathrm{nuc}$}    \\
\hline
He & 2  & 3.8[-9] & 2.4[-22] \\
Ne & 10 & 2.2[-6] & 1.5[-19] \\
Ar & 18 & 7.4[-5] & 5.2[-18] \\
Kr & 36 & 3.6[-3] & 3.1[-16] \\
Xe & 54 & 4.5[-2] & 5.3[-15] \\
Rn & 86 & 1.07    & 2.2[-13] \\
\hline\hline
\end{tabular}
\caption{CP-violating polarizability, $\beta^\mathrm{CP}$, in
Gaussian atomic units, for rare-gas atoms. CP-violation is either
due to the electron EDM, $d_e$, or due to the neutral currents \eref{Eq:NC}.
Notation $x[y]$ stands for $x \times 10^y$.
\label{table:betaValues} }
\end{table}
\end{center}

\begin{figure}[ht]
\includegraphics[scale= 0.7,origin= c,draft= false]{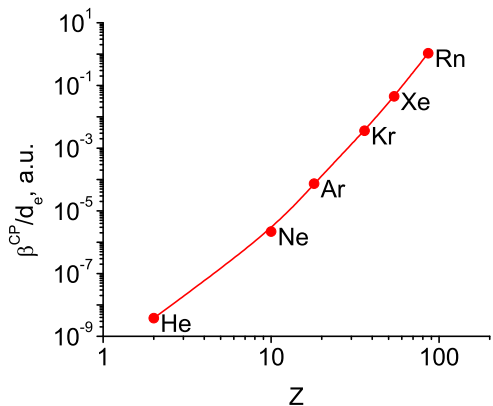}
\caption{ (Color online)
Dependence of the  CP-violating polarizability $\beta^\mathrm{CP}$ on the nuclear
charge $Z$ for rare-gas atoms.
CP-violation is due to the electron EDM, $d_e$.
The ratio $\beta^\mathrm{CP}/d_e$ is given in atomic units.}
\label{Fig:betaPlot}
\end{figure}

Numerical results for rare-gas atoms are presented in
Table~\ref{table:betaValues} and also plotted in
Fig.~\ref{Fig:betaPlot}.  In Table~\ref{table:betaValues}, the values in the
column marked $\beta^\mathrm{CP}/d_e$ were computed directly,
while the values $\beta^\mathrm{CP}/k_1^\mathrm{nuc}$
(the last column) were obtained from $\beta^\mathrm{CP}/d_e$ as
explained in Section~\ref{Sec:Z5scaling}.

From Fig.~\ref{Fig:betaPlot} we observe a pronounced dependence of the
values on the nuclear charge $Z$. Such a steep scaling of the
CP-odd polarizabilities is expected from the considerations
presented in \sref{Sec:Z5scaling}.

To illustrate the (doubly) relativistic origin of the CP-odd
polarizability $\beta^\mathrm{CP}$, we compile values of various
contributions to $\beta^\mathrm{CP}$ in Table~\ref{table:break}
for an isolated Xe atom. Apparently, the dominant contributions
are from $\langle \mu^\mathrm{CP}\rangle_{1,a}$,
Eq.~(\ref{Eq:muDHF1a}), and $\langle
\mu^\mathrm{CP}\rangle_{1,b}$, Eq.~(\ref{Eq:muDHF1b}), but there
is strong cancelation between these two terms. As we will see
below, this cancelation is not accidental.


\begin{center}
\begin{table}[h]
\begin{tabular}{lddd}
\hline\hline \multicolumn{1}{c}{$k$} &
\multicolumn{1}{c}{$\beta^\mathrm{CP}_{k,a}/d_{e}$} &
\multicolumn{1}{c}{$\beta^\mathrm{CP}_{k,b}/d_{e}$} &
\multicolumn{1}{c}{sum}   \\
\hline
1& -0.108 & 0.132 & 2.44[-2]  \\
2& 6.53[-3] & -6.63[-5] & 6.46[-3] \\
3 & 8.19[-3] & 5.13[-3] & 1.33[-2] \\
\hline
total & & & 4.42[-2] \\
\hline\hline
\end{tabular}
\caption{ Contributions to CP-violating polarizability,
$\beta^\mathrm{CP}/d_{e}$, in Gaussian
atomic units, for an isolated Xe atom.
Each contribution is defined via
Eq.~(\protect\ref{Eq:muDHF1a})--(\protect\ref{Eq:muDHF3b}) as
$\beta_{k,\alpha}^\mathrm{CP}/d_{e}=  \langle
\mu^\mathrm{CP}\rangle_{k,\alpha}/(d_{e}\mathcal{E}_0)$.
CP-violation is due to the electron EDM, $d_e$.
Notation $x[y]$ stands for $x \times 10^y$.\label{table:break} }
\end{table}
\end{center}

\subsection{$Z^5$ scaling and relation between EDM and NC contributions}
\label{Sec:Z5scaling}

Let us consider non-relativistic limit of
Eqs.~(\ref{Eq:muBreakPT1} --~\ref{Eq:muBreakPT3}). The magnetic
moment operator is reduced to the form:
\begin{equation}\label{Eq:Zsc1}
\bm{\mu}=\frac{\alpha}{2}(2\bm{s}+\bm{l}).
\end{equation}
This operator can not change electronic principal quantum numbers.
Because of that the contributions \eref{Eq:muBreakPT2} and
\eref{Eq:muBreakPT3} vanish, as there $\bm{\mu}$ should mix
occupied and excited orbitals. Thus, we are left with the single
term \eref{Eq:muBreakPT1}, which can be further split in two parts
\eref{Eq:muDHF1a} and \eref{Eq:muDHF1b}. We will show now that
these two parts cancel each other.

Indeed, in the non-relativistic approximation the operator $V^\mathrm{CP}$
is given by a scalar product of the spin vector and the orbital
vector. Therefore, in the $LS$-coupling scheme it can couple the
ground state $^1\!S_0$ only with excited states $^3\!P_0$.
Operator \eref{Eq:Zsc1} is diagonal in the quantum numbers $L$ and
$S$ and can couple $^3\!P_0$ only with $^3\!P_1$. To return back
to the ground state, the dipole operator $V^\mathrm{ext}$ has to
couple $^3\!P_1$ with $^1\!S_0$. However, this matrix element
vanishes in the non-relativistic approximation. The above states
$^3\!P_{0,1}$ are formed from the excited electron and a whole in
the core, which correspond to two expressions \eref{Eq:muDHF1a}
and \eref{Eq:muDHF1b}. We conclude that these two contributions
exactly cancel in the non-relativistic approximation.

The matrix element $\langle ^3\!P_1| V^\mathrm{ext} |^1\!S_0
\rangle$ is proportional to the spin-orbit mixing, which is of the
order of $(\alpha Z)^2$. It follows from \eref{Eq:mu_ME} that
relativistic correction to operator \eref{Eq:Zsc1} is of the same
order. This correction accounts for the nondiagonal in the
principle quantum numbers matrix elements of $\bm{\mu}$ and leads
to the nonzero values of the terms \eref{Eq:muBreakPT2} and
\eref{Eq:muBreakPT3}. Thus, we see that all three terms in
\eref{Eq:muBreakPT} are suppressed by the relativistic factor
$(\alpha Z)^2$, in agreement with numerical results from
\tref{table:break}.

Matrix elements of the CP-odd interaction $V^\mathrm{CP}$ depend
on the short distances and rapidly decrease with quantum number
$j$. To a good approximation it is possible to neglect all matrix
elements for $j\ge 3/2$. For the remaining matrix elements between
orbitals $s_{1/2}$ and $p_{1/2}$ an analytical expression can be found
in \cite{Khr91}:
\begin{align}
\langle s_{1/2} |V^\mathrm{CP\!,EDM}|p_{1/2}\rangle
&=\frac{16}{3}\frac{\alpha^2 Z^3 R^\mathrm{EDM}}{(\nu_s\nu_p)^{3/2}}d_e,
\label{Eq:Zsc2}\\
\langle s_{1/2} |V^\mathrm{CP\!,NC}|p_{1/2}\rangle
&=\frac{G_\textrm{F}}{2\sqrt{2}\pi}
\frac{\alpha Z^3 R^\mathrm{NC}}{(\nu_s\nu_p)^{3/2}}k_1^\mathrm{nuc},
\label{Eq:Zsc3}
\end{align}
where we use effective quantum numbers $\nu=(-2\veps)^{-1/2}$.
$R^\mathrm{EDM}$ and $R^\mathrm{NC}$ are relativistic enhancement
factors:
\begin{align}
R^\mathrm{EDM} &= \frac{3}{\gamma(4\gamma^2-1)}
=\left\{
\begin{array}{lll}
1,   & Z=1,\\
1.4, & Z=54, & \mathrm{(Xe)},\\
2.7, & Z=86, & \mathrm{(Rn)},
\end{array}
\right.
\label{Eq:Zsc4}\\
R^\mathrm{NC} &=
\frac{4\gamma(2Zr_\textrm{N})^{2\gamma-2}}{\Gamma^2(2\gamma+1)}
=\left\{
\begin{array}{ll}
1,   & Z=1,\\
2.5, & Z=54,\\
8.7, & Z=86,
\end{array}
\right.
\label{Eq:Zsc5}
\end{align}
where $\gamma=\sqrt{1-(\alpha Z)^2}$ and the radius of the nucleus is taken to be
$r_\textrm{N}=1.2\,(Z+N)^{1/3} \mathrm{fm}$ \cite{Khr91}.

We see that both CP-odd operators scale as $Z^3R$ with
relativistic enhancement factors $R$ given by \eref{Eq:Zsc4} and
\eref{Eq:Zsc5}. This scaling adds up with relativistic suppression
$(\alpha Z)^2$ discussed above to give overall scaling $Z^5 R$.
This scaling agrees with our numerical calculations and \fref{Fig:betaPlot}.

Because of the similarity between matrix elements \eref{Eq:Zsc2}
and \eref{Eq:Zsc3} of operators $V^\mathrm{CP\!,EDM}$ and
$V^\mathrm{CP\!,NC}$, there is no need in calculating
independently the NC contribution to $\beta^\textrm{CP}$. It is
sufficient to substitute matrix elements \eref{Eq:Zsc2} in all
equations with matrix elements \eref{Eq:Zsc3}. Comparing these
expressions we find that to get the contribution to
$\beta^\textrm{CP}$ induced by the CP-odd weak neutral currents we
need to make following substitution:
\begin{equation}\label{Eq:Zsc6}
\frac{d_e}{e r_0} \Longleftrightarrow 0.64\times 10^{-13}
\frac{R^\mathrm{NC}}{R^\mathrm{EDM}}k_1^\mathrm{nuc},
\end{equation}
where $r_0$ is the Bohr radius and $R^\mathrm{EDM}$ and
$R^\mathrm{NC}$ are given by \eref{Eq:Zsc4} and \eref{Eq:Zsc5}.
The accuracy of \Eref{Eq:Zsc6} is typically 15~--~20\%, which is
sufficient for our purposes. It was used to calculate the last
column of \tref{table:betaValues}.

\section{Limits on electron EDM from measurement of CP-odd polarizability}
\label{Sec:LXe}

\begin{figure}[ht]
\includegraphics[scale= 0.5]{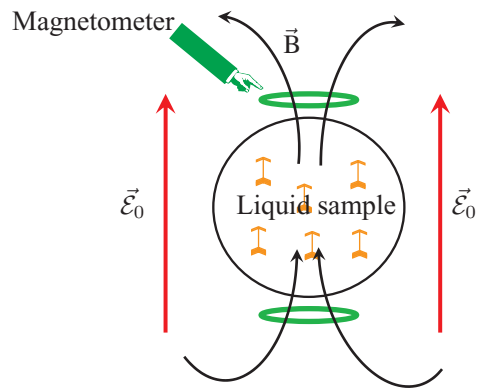}
\caption{(Color online) A scheme for measuring CP-violating polarizability.}
\label{Fig:Setup}
\end{figure}

Here we envision the following experimental setup (see
Fig.~\ref{Fig:Setup}) to
measure the CP-violating polarizability:
A strong electric field $\mathcal{E}_0$ is applied to a sample
of diamagnetic atoms of number density $n$. A macroscopic magnetization
arises due to the CP-violating polarizability. This magnetization
generates
a very weak magnetic field $B$. One could measure this induced magnetic
field and set the limits on the electron EDM or other CP-violating
mechanisms.
In particular, for a spherical cell the maximum value of the generated
magnetic
field at the surface of the sphere can be related to the CP-violating
polarizability as
\begin{equation}
B_{\max}= \frac{8\pi}{3} n \, \beta^\mathrm{CP} \mathcal{E}_0 \, .
\label{Eq:Signal}
\end{equation}
Clearly, one should increase the number density to enhance the signal,
and it is beneficial to work with a dense liquid or solid sample.

Among the rare-gas atoms, considered here, xenon
has the most suitable properties for such an experiment: Xe is the
heaviest
non-radioactive atom, it has a large number density ($n \sim
10^{22} \, 1/\mathrm{cm}^{3}$), and liquid Xe  has a
high electric field breakdown strength ($\mathcal{E}_0 \sim
4 \times 10^5 \mathrm{V/cm}$). Our calculations in
Section~\ref{Sec:Results} were carried out
for isolated atoms. However, in a liquid, there are certain
environmental effects (such as confinement of electronic density) that
affect
the CP-violating signal.
To estimate the confinement effects in the liquid, we employ
the liquid-cell model. The calculations are similar to those presented
in Ref.~\cite{RavDer04}.
In brief, we solve the DHF equations for a Xe atom
in a spherical cavity of radius $R_\mathrm{cav} =
\left(\frac{3}{4\pi}\, \frac{1}{n}  \right)^{1/3}$,
with certain boundary conditions
imposed at the cavity surface.
For a density
of LXe of 500 amagat~\cite{AmagatDef},
$R_\mathrm{cav} \simeq 4.9 $ bohr. For a solid state,
$R_\mathrm{cav} \simeq 4.2 $ bohr and we use the latter
in the calculations (see discussion in Ref.~\cite{RavDer04}).
Technically, we applied the variational Galerkin method on a set
of 100 B-spline functions~\cite{JohBluSap88}.
We find numerically that compared to an isolated atom, the CP-violating
polarizability of a Xe atom in LXe is reduced by about 65\%,
\begin{equation}
  \beta^\mathrm{CP} (\mathrm{LXe}) \approx 1.5 \times 10^{-2} d_e \, .
  \label{Eq:betaLXe}
\end{equation}

From Eq.~(\ref{Eq:Signal}) it is clear that the more sensitive the
measurement of the B-field,
the tighter the constraints on $\beta^\mathrm{CP}$ (and $d_e$) are.
Presently, the most sensitive measurement
of weak magnetic fields has been carried out by Princeton
group~\cite{KomKorAll03}. Using atomic
magnetometry, this
group has reached the sensitivity level of $5.4 \times 10^{-12} \,
\mathrm{G}/\sqrt{\mathrm{Hz}}$.
 The projected theoretical limit~\cite{KomKorAll03} of this method is
$10^{-13} \, \mathrm{G}/\sqrt{\mathrm{Hz}}$. Notice that this
estimate has been carried out for a sample of volume 0.3 cm$^3$.
According to Romalis~\cite{RomPrivate},  the sensitivity increases with volume $V$ as
$V^{1/3}$,
so a 100 cm$^3$ cell would have an even better sensitivity
of about $10^{-14} \, \mathrm{G}/\mathrm{Hz}^{1/2}$.
More optimistic estimate, based on
nonlinear Faraday effect in atomic vapors~\cite{BudKimRoc00}, is given
in Ref.\cite{Lam02}; here the projected sensitivity is
$3 \times 10^{-15} \, \mathrm{G}/\sqrt{\mathrm{Hz}}$.

Assuming 10 days of averaging,
the most optimistic published estimate of the sensitivity to magnetic
field~\cite{Lam02}
leads to the weakest measurable field of $B \simeq 3 \times 10^{-18} \,
\mathrm{G}$.
Combining this estimate with
the breakdown strength of the E-field for LXe, $\mathcal{E}_0 \sim
4 \times 10^5 \, \mathrm{V/cm}$, and our computed value of CP-odd
polarizability, Eq.~(\ref{Eq:betaLXe}),
we arrive at the constraint on the electron EDM,
\begin{equation}
d_{e}(\mathrm{LXe}) < 6 \times 10^{-26} \, e \cdot \mathrm{cm}.
\end{equation}
This projected limit is more than an order of magnitude worse than the present limit on the
electron EDM from the Tl experiment~\cite{RegComSch02},
$d_{e}\left( \mathrm{Tl} \right)  <1.6 \times 10^{-27}\, e \cdot
\mathrm{cm}$. It is worth emphasizing that the above limit has
been obtained using B-field sensitivity estimate from
Ref.~\cite{Lam02}; with the present sensitivity
record~\cite{KomKorAll03}, the constraints of electron EDM are
several orders of magnitude weaker. In other words, we find that a
substantial improvement in the experimental sensitivity to weak
magnetic fields is required before the CP-violating polarizability
of LXe can be used for EDM searches.

\section{Conclusion}
\label{Sec:Conclusion}
To summarize, we have computed novel CP-violating
atomic polarizabilities~\cite{Bar04}, $\beta^\mathrm{CP}$, for rare-gas atoms.
We have derived third-order expressions for $\beta^\mathrm{CP}$ and
employed the Dirac-Hartree-Fock method to evaluate the resulting
expressions. We have elucidated the doubly relativistic origin
of the polarizability and demonstrated strong $Z^5$ dependence
on the nuclear charge. Finally, we
evaluated a feasibility of setting a limit on the electron
EDM by measuring CP-violating magnetization
of liquid Xe.
We found that such an experiment could provide competitive bounds on electron EDM only
if the present level of experimental sensitivity to ultra-weak magnetic fields~\cite{KomKorAll03}
is improved by several orders of magnitude.

\acknowledgments

We would like to thank M. Romalis for motivating discussion and overview of
experimental techniques,
D. Budker for a communication regarding the weakest measurable magnetic fields,
and S. G. Porsev for discussions. This
work was supported in part by the  NSF
Grant No. PHY-0354876, NIST precision measurement grant, Russian
RFBR grant 05-02-16914, and
by the NSF through a grant to the Institute for Theoretical Atomic,
Molecular, and Optical Physics at Harvard University and the Smithsonian
Astrophysical Observatory.

\bibliographystyle{apsrev}


\begin{thebibliography}{21}
\expandafter\ifx\csname natexlab\endcsname\relax\def\natexlab#1{#1}\fi
\expandafter\ifx\csname bibnamefont\endcsname\relax
  \def\bibnamefont#1{#1}\fi
\expandafter\ifx\csname bibfnamefont\endcsname\relax
  \def\bibfnamefont#1{#1}\fi
\expandafter\ifx\csname citenamefont\endcsname\relax
  \def\citenamefont#1{#1}\fi
\expandafter\ifx\csname url\endcsname\relax
  \def\url#1{\texttt{#1}}\fi
\expandafter\ifx\csname urlprefix\endcsname\relax\def\urlprefix{URL }\fi
\providecommand{\bibinfo}[2]{#2}
\providecommand{\eprint}[2][]{\url{#2}}

\bibitem[{\citenamefont{Khriplovich and Lamoreaux}(1997)}]{KL97}
\bibinfo{author}{\bibfnamefont{I.~B.} \bibnamefont{Khriplovich}}
  \bibnamefont{and} \bibinfo{author}{\bibfnamefont{S.~K.}
  \bibnamefont{Lamoreaux}}, \emph{\bibinfo{title}{{CP} Violation without
  Strangeness}} (\bibinfo{publisher}{Springer}, \bibinfo{address}{Berlin},
  \bibinfo{year}{1997}).

\bibitem[{\citenamefont{Bigi and Sanda}(2000)}]{BS00}
\bibinfo{author}{\bibfnamefont{I.~I.} \bibnamefont{Bigi}} \bibnamefont{and}
  \bibinfo{author}{\bibfnamefont{A.~I.} \bibnamefont{Sanda}},
  \emph{\bibinfo{title}{CP Violation}} (\bibinfo{publisher}{Cambridge
  University Press}, \bibinfo{address}{Cambridge}, \bibinfo{year}{2000}).

\bibitem[{\citenamefont{Baryshevsky}(2004)}]{Bar04}
\bibinfo{author}{\bibfnamefont{V.~G.} \bibnamefont{Baryshevsky}},
  \bibinfo{journal}{Phys. Rev. Lett.} \textbf{\bibinfo{volume}{93}},
  \bibinfo{pages}{043003} (\bibinfo{year}{2004}).

\bibitem[{\citenamefont{Shapiro}(1968)}]{Sha68}
\bibinfo{author}{\bibfnamefont{F.~L.} \bibnamefont{Shapiro}},
  \bibinfo{journal}{Sov. Phys. Uspekhi} \textbf{\bibinfo{volume}{11}},
  \bibinfo{pages}{345} (\bibinfo{year}{1968}).

\bibitem[{\citenamefont{Vasiliev and Kolycheva}(1978)}]{VK78}
\bibinfo{author}{\bibfnamefont{B.~V.} \bibnamefont{Vasiliev}} \bibnamefont{and}
  \bibinfo{author}{\bibfnamefont{E.~V.} \bibnamefont{Kolycheva}},
  \bibinfo{journal}{Sov. Phys.--JETP} \textbf{\bibinfo{volume}{47}},
  \bibinfo{pages}{243} (\bibinfo{year}{1978}).

\bibitem[{\citenamefont{Lamoreaux}(2001)}]{Lam01}
\bibinfo{author}{\bibfnamefont{S.~K.} \bibnamefont{Lamoreaux}},
  \bibinfo{journal}{Phys. Rev. A} \textbf{\bibinfo{volume}{66}},
  \bibinfo{pages}{022109} (\bibinfo{year}{2001}).

\bibitem[{\citenamefont{Regan et~al.}(2002)\citenamefont{Regan, Commins,
  Schmidt, and DeMille}}]{RegComSch02}
\bibinfo{author}{\bibfnamefont{B.~C.} \bibnamefont{Regan}},
  \bibinfo{author}{\bibfnamefont{E.~D.} \bibnamefont{Commins}},
  \bibinfo{author}{\bibfnamefont{C.~J.} \bibnamefont{Schmidt}},
  \bibnamefont{and} \bibinfo{author}{\bibfnamefont{D.}~\bibnamefont{DeMille}},
  \bibinfo{journal}{Phys. Rev. Lett.} \textbf{\bibinfo{volume}{88}},
  \bibinfo{pages}{071805} (\bibinfo{year}{2002}).

\bibitem[{\citenamefont{Hunter}(2004)}]{Hun04}
\bibinfo{author}{\bibfnamefont{L.}~\bibnamefont{Hunter}}
  (\bibinfo{year}{2004}), \bibinfo{note}{book of abstracts of {DAMOP} meeting}.

\bibitem[{\citenamefont{Ravaine and Derevianko}(2004{\natexlab{a}})}]{RD04}
\bibinfo{author}{\bibfnamefont{B.}~\bibnamefont{Ravaine}} \bibnamefont{and}
  \bibinfo{author}{\bibfnamefont{A.}~\bibnamefont{Derevianko}},
  \bibinfo{journal}{Phys. Rev. A} \textbf{\bibinfo{volume}{69}},
  \bibinfo{pages}{050101(R)} (\bibinfo{year}{2004}{\natexlab{a}}).

\bibitem[{\citenamefont{Sandars}(1965)}]{San65}
\bibinfo{author}{\bibfnamefont{P.~G.~H.} \bibnamefont{Sandars}},
  \bibinfo{journal}{Phys. Lett.} \textbf{\bibinfo{volume}{14}},
  \bibinfo{pages}{194} (\bibinfo{year}{1965}).

\bibitem[{\citenamefont{Sandars}(1966)}]{San66}
\bibinfo{author}{\bibfnamefont{P.~G.~H.} \bibnamefont{Sandars}},
  \bibinfo{journal}{Phys. Lett.} \textbf{\bibinfo{volume}{22}},
  \bibinfo{pages}{290} (\bibinfo{year}{1966}).

\bibitem[{\citenamefont{Jungmann}(2005)}]{Jun05}
\bibinfo{author}{\bibfnamefont{K.~P.} \bibnamefont{Jungmann}}
  (\bibinfo{year}{2005}), \bibinfo{note}{\eprint{physics/0501154}}.

\bibitem[{\citenamefont{Khriplovich}(1991)}]{Khr91}
\bibinfo{author}{\bibfnamefont{I.~B.} \bibnamefont{Khriplovich}},
  \emph{\bibinfo{title}{Parity non-conservation in atomic phenomena}}
  (\bibinfo{publisher}{Gordon and Breach}, \bibinfo{address}{New York},
  \bibinfo{year}{1991}).

\bibitem[{\citenamefont{Landau and Lifshitz}(1997)}]{LanLif97}
\bibinfo{author}{\bibfnamefont{L.~D.} \bibnamefont{Landau}} \bibnamefont{and}
  \bibinfo{author}{\bibfnamefont{E.~M.} \bibnamefont{Lifshitz}},
  \emph{\bibinfo{title}{Quantum Mechanics}}, vol. \bibinfo{volume}{III}
  (\bibinfo{publisher}{Butterworth-Heinemann}, \bibinfo{year}{1997}),
  \bibinfo{edition}{3rd} ed.

\bibitem[{\citenamefont{Johnson et~al.}(1988)\citenamefont{Johnson, Blundell,
  and Sapirstein}}]{JohBluSap88}
\bibinfo{author}{\bibfnamefont{W.~R.} \bibnamefont{Johnson}},
  \bibinfo{author}{\bibfnamefont{S.~A.} \bibnamefont{Blundell}},
  \bibnamefont{and}
  \bibinfo{author}{\bibfnamefont{J.}~\bibnamefont{Sapirstein}},
  \bibinfo{journal}{Phys.\ Rev.\ A} \textbf{\bibinfo{volume}{37}},
  \bibinfo{pages}{307} (\bibinfo{year}{1988}).

\bibitem[{\citenamefont{Ravaine and Derevianko}(2004{\natexlab{b}})}]{RavDer04}
\bibinfo{author}{\bibfnamefont{B.}~\bibnamefont{Ravaine}} \bibnamefont{and}
  \bibinfo{author}{\bibfnamefont{A.}~\bibnamefont{Derevianko}},
  \bibinfo{journal}{Phys. Rev. A} \textbf{\bibinfo{volume}{69}},
  \bibinfo{pages}{050101(R)} (\bibinfo{year}{2004}{\natexlab{b}}).

\bibitem[{Ama()}]{AmagatDef}
\bibinfo{note}{Amagat density unit is equal to 44.615 moles per cubic meter
  (mol/m$^3$)}.

\bibitem[{\citenamefont{Kominis et~al.}(2003)\citenamefont{Kominis, Kornack,
  Allred, and Romalis}}]{KomKorAll03}
\bibinfo{author}{\bibfnamefont{I.~K.} \bibnamefont{Kominis}},
  \bibinfo{author}{\bibfnamefont{T.~W.} \bibnamefont{Kornack}},
  \bibinfo{author}{\bibfnamefont{J.~C.} \bibnamefont{Allred}},
  \bibnamefont{and} \bibinfo{author}{\bibfnamefont{M.~V.}
  \bibnamefont{Romalis}}, \bibinfo{journal}{Nature}
  \textbf{\bibinfo{volume}{422}}, \bibinfo{pages}{596} (\bibinfo{year}{2003}).

\bibitem[{Rom()}]{RomPrivate}
\bibinfo{note}{{M.} Romalis, private communications}.

\bibitem[{\citenamefont{Budker et~al.}(2000)\citenamefont{Budker, Kimball,
  Rochester, Yashchuk, and Zolotorev}}]{BudKimRoc00}
\bibinfo{author}{\bibfnamefont{D.}~\bibnamefont{Budker}},
  \bibinfo{author}{\bibfnamefont{D.~F.} \bibnamefont{Kimball}},
  \bibinfo{author}{\bibfnamefont{S.~M.} \bibnamefont{Rochester}},
  \bibinfo{author}{\bibfnamefont{V.~V.} \bibnamefont{Yashchuk}},
  \bibnamefont{and}
  \bibinfo{author}{\bibfnamefont{M.}~\bibnamefont{Zolotorev}},
  \bibinfo{journal}{Phys. Rev. A} \textbf{\bibinfo{volume}{62}},
  \bibinfo{pages}{043403} (\bibinfo{year}{2000}).

\bibitem[{\citenamefont{Lamoreaux}(2002)}]{Lam02}
\bibinfo{author}{\bibfnamefont{S.~K.} \bibnamefont{Lamoreaux}},
  \bibinfo{journal}{Phys. Rev. A} \textbf{\bibinfo{volume}{66}},
  \bibinfo{pages}{022109} (\bibinfo{year}{2002}).

\end{thebibliography}

\end{document}